# CHOIR TRANSFORMER: GENERATING POLYPHONIC MUSIC WITH RELATIVE ATTENTION ON TRANSFORMER

*Jiuyang Zhou*     *Hong Zhu\**     *Xingping Wang*

School of Automation and Information Engineering, Xi'an University of Technology

**ABSTRACT**

Polyphonic music generation is still a challenge direction due to its correct between generating melody and harmony. Most of the previous studies used RNN-based models. However, the RNN-based models are hard to establish the relationship between long-distance notes. In this paper, we propose a polyphonic music generation neural network named Choir Transformer[1], with relative positional attention to better model the structure of music. We also proposed a music representation suitable for polyphonic music generation. The performance of Choir Transformer surpasses the previous state-of-the-art accuracy of 4.06%. We also measures the harmony metrics of polyphonic music. Experiments show that the harmony metrics are close to the music of Bach. In practical application, the generated melody and rhythm can be adjusted according to the specified input, with different styles of music like folk music or pop music and so on.

*Index Terms*— Music language model, polyphonic music generation, symbolic music generation, Transformer, relative positional attention

## 1. INTRODUCTION

Music is composed of notes in a specific relationship. The horizontal dependencies and vertical dependencies in polyphonic music constitute music texture. Music has rich semantic properties like motif, verse, chorus, and bridge. Due to the different textures of music, music is divided into monophonic, homophonic, and polyphonic[1]. Monophonic music has only one melody without accompanying melodies. A central theme dominates homophonic music; other melodies are used as harmony or rhythmic collaboration, while polyphonic music comprises several independent melodies. According to the semantic features of music texture, the main feature of monophonic music is the horizontal dependencies.

In polyphonic music, the horizontal dependencies form the melody, and the vertical dependencies form the chord, as shown in Fig.1. The combination will only constitute harmony if one generates the right out-of-chord tone.

This paper mainly studies the symbolic music generation of polyphonic music. We selected Bach's choral hymns for polyphonic music model training. The polyphonic generation task needs to learn the melody features while simultaneously paying attention to the vertical dependencies' harmony. We propose a data representation method suitable for polyphonic music generation and two data enhancement methods to design the input data representation method.

Polyphonic music modeling has always been challenging in the direction of polyphonic music generation. Two difficulties are the main direction of exploration: how to make the input information fully contain the information of polyphonic music, and the other is to choose a model to generate well. The current semantic model is mainly based on RNN-based[2] models. For example, DeepBach[3] uses DeepRNN[4] for feature extraction; the TonicNet[5] model uses chord and pitch information as a data representation method, using the GRU[6] model for note encoding and feature extraction; DeepChoir[7] uses Bi-LSTM[8] for sequence modeling to construct a stacked encoder-decoder structure model.

In general, previous works based on RNNs[9][10] reproduce the style of Bach quite well and also improve the performance of generating polyphonic music by adding extra information. However, as the generation sequence keeps growing in practical applications, the model has difficulty catching global and local information, causing a drop in the quality of generated samples. Although improved models have been working to solve the long-term dependency problem, taking LSTM as an example, the model can use about 200 context tokens. In contrast, the distant past is modeled only as a rough semantic field or topic[11].

This paper starts with a Transformer[12] baseline for polyphonic music modeling, infusing relative distance in attention to help improve the semantic feature extraction of music. In the experiment, we achieve state-of-the-art performance. In order to measure the texture of polyphonic music, this paper measures the melody and chord in the texture. The experiments showed that our model is close to the metrics of the music of Bach.

Our main contributions are to：

---

[1] https://github.com/Zjy0401/choir-transformer

- We propose Choir Transformer, A Transformer based polyphonic generation model, achieved higher accuracy on validation than previous models.
- For the polyphonic texture, we propose a data representation method suitable for polyphonic music, and we propose a relative position attention mechanism to improve the generated quality and generate a longer effective length.

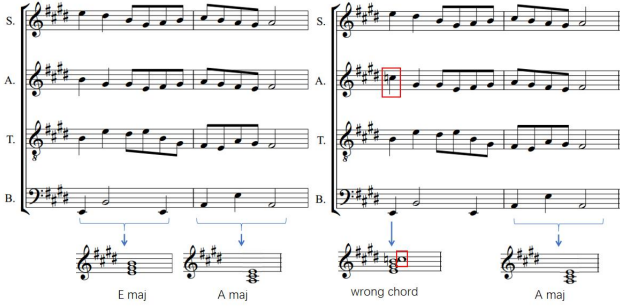

**Fig.1** The horizontal dependencies are notes from left to right, and the vertical dependencies are notes from top to bottom which form the chord. If a note in any part is wrong, the wrong harmony will be generated, as shown in the correct figure in the Red box

## 2. METHODOLOGY

### 2.1 Music Representation

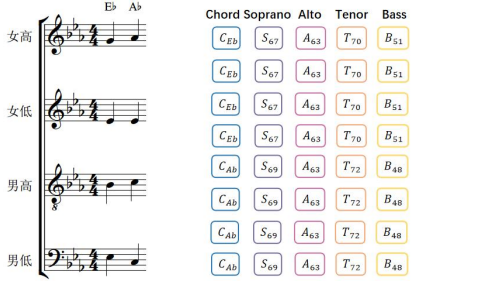

a) Choral music excerpt    b) our data representation

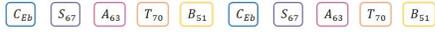

c) input data format

**Fig.2** Data representation process from choral music to input, the time resolution is set to 16th notes.

We build two types of events: Note events represent the pitch of a note, with 129 different numbers, including 128 pitch tokens and one rest token. Chord events, represent the chord tokens, with 12 major, minor, augmented and diminished chords, and one other chord token, for a total of 49 types. We arrange the polyphonic music data to combine information in a chord-first manner.

Formally, This paper give the sequence of input $x = \{C_{cls}^1, S_{note}^1, A_{note}^1, T_{note}^1, B_{note}^1, \ldots, C_{cls}^{end}, S_{note}^{end}, A_{note}^{end}, T_{note}^{end}, B_{note}^{end}\}$

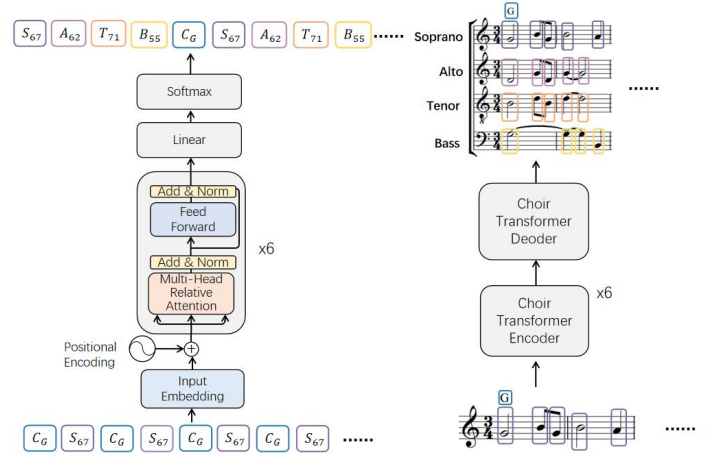

**Fig.3** The architecture of the Choir Transformer. When generating, the input data is combined with the chord and soprano melody, and the output is combined with the chord and four-part melody. When training, the input data is combined with chord and four-part melody

The pitches of four parts are expressed as $S_{note}^t, A_{note}^t, T_{note}^t, B_{note}^t$, chord as $C_{cls}^t$, t represents time, $t \in [1, end]$, cls is the classes of chord.

### 2.2 Choir Transformer

Choir Transformer is a multi-head autoregressive encoder-decoder model, as shown in Fig.5. The encoder uses an improved Transformer decoder, and the decoder uses a Linear layer and softmax. We combined the note information and chord information as input sequences. The inputs are represented by word embedding and positional encoding as $X \in R^{T_p \times d_p}$, $T_{(\cdot)}$ and $d_{(\cdot)}$ represent sequence length and feature dimension respectively, as Equation(2):

$$X = \text{word embedding}(x) + \text{positional encoding}(x) \quad (2)$$

Define $Q_i, K_i, V_i$ as attention vector and calculate as Equation(3):

$$Q_i = XW^{Q_i}, K_i = XW^{K_i}, V_i = XW^{V_i} \quad (3)$$

Where $W^Q \in R^{d_x \times d_Q}, W^K \in R^{d_x \times d_Q}, W^V \in R^{d_x \times d_Q}$ are each square matrices of attention weights. Towards multi-relative attention as follows Equations(4)(5):

$$head_i = \text{Softmax}(\frac{Q_i K_i^T + Q_i R_i^T}{\sqrt{D}})V_i \quad (4)$$

$$Z = \text{Concat}(head_1, \ldots, head_i)W^O \quad (5)$$

The attention outputs for each head are concatenated and linearly transformed to get Z. A upper triangular mask ensures that queries cannot attend to keys later in the sequence. The feedforward sub-layer takes the output Z from the previous attention sub-layer, and performs two layers of point-wise layers on the depth D dimension, as shown in Equation(6). $W_1, W_2, b_1, b_2$ are weights and biases of those two layers.

$$FF(Z) = \text{ReLU}(ZW_1 + b_1)W_2 + b_2 \quad (6)$$

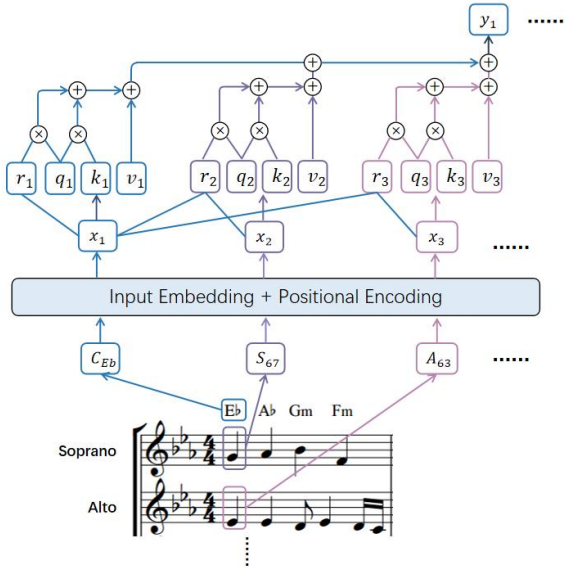

**Fig.4** The Relative-attention of Choir Transformer, the prediction of next token will calculate attention by adding the relative position information

### 2.3 Relative Positional Attention

The positional encoding in Transformer is absolute. However, the relative position plays an essential role in music structure.

In Choir Transformer, inspired by Music Transformer[13], we propose the relative positional attention, the detail shows on Fig.4. Through model training, the model learned the relationship between relative positions R, so that the distance between two positions establishes the position in the sequence. We calculate the relative position offset with R and Q as follows:

$$Z = \text{Softmax}(\frac{QK^T + QR^T}{\sqrt{D}})V \tag{8}$$

## 3. EXPERIMENTS

### 3.1 Dataset

Based on the JS Bach Chorales[14], we selected the JS Fake Chorales dataset[15] and expanded similar music by algorithm. JS Fake Chorales dataset has 500 four-part chorus music and added chord marks manually.

We propose two data enhancements, the transposition and inversion methods. The transposition method can transpose to 11 tones, then inverse the music. Through the two enhancements, we expand the dataset to 12000 chorus music. The melody can be shifted up and down with different tones, so the interval relationship is essential. The model learns the interval relationship on a larger dataset. In order to further explore the interval relationship, we also add the reverse sequences for training.

### 3.2 Quantitative Evaluations

#### 3.2.1 Ablation experiment

We performed the ablation on Choir Transformer, using the accuracy of the validation set as the measurement. Shown in Table 1. (Chord: added chord on data representation, rpr: added relative-attention, amp: using transposition to augment, rev: using reverse method to augment).

| Chord | rpr | amp | rev | Accuracy(%) |
|-------|-----|-----|-----|-------------|
|       |     |     |     | 86.22 |
| √     |     |     |     | 88.84 |
| √     | √   |     |     | 89.39 |
| √     | √   | √   |     | 92.33 |
| √     | √   | √   | √   | 94.99 |

Table 1 Validation accuracy of ablation experiments

Experiments show that using relative positional attention and adding chord information improves the model's performance. Through two dataset enhancement methods, the accuracy of the validation set is optimal.

Afterward, we measure the harmony of polyphonic music. We selected the Bach choral music in the test set to compare the metrics of the generated music. The measure of the experiment shows that the generated samples are very close to Bach's works on harmony.

#### 3.2.2 Harmonic Metrics and Evaluations

From the point of harmonic, the statistical metrics of chords and notes provide the metrics in polyphonic music. We choose three harmony metrics of chords and notes as texture metrics, which are proposed in[16]. The metrics have been extensively used in the melody harmonization task[17][18][19].

- Chord Tone to non-Chord tone Ratio(CTnCTR): calculates the ratio of the number of chord notes to the number of the non-chord notes.
- Pitch Consonance Score(PCS): Calculated based on the interval between melody and chord, given different weights based on different intervals.
- Melody-Chord Tonal Distance(MCTD): Calculated the similarity between the note feature and chord vector[20] in 6-D space[21].

#### 3.2.3 Comparative Test

We selected the recently released DeepChoir, TonicNet, and DeepBach models for comparative experiments, as shown in Figure 6. We use Token Error Rate(%) as a metric to measure the modeling ability of the model. The input of RNNs differs from the Choir Transformer, so we set the output of the soprano to 0 and do not participate in the generated comparison.

Experiments show that Choir Transformer surpasses the previous state-of-the-art accuracy.

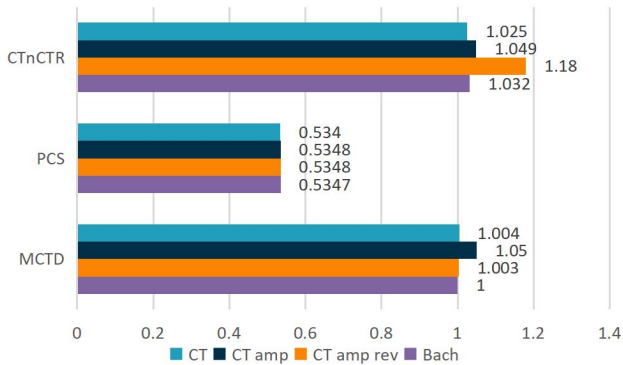

**Fig.5** Comparison of different Choir Transformer and Bach's original song indicators, CT is the Choir Transformer baseline, amp is trained using the transposition method, rev uses the reverse music method.

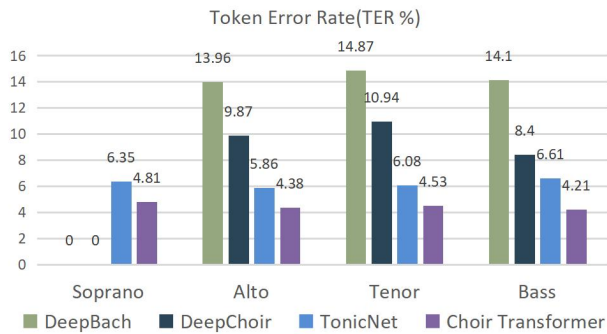

**Fig.6** Comparison of Token error rate (TER) indicators.

### 3.3 Qualitative Evaluation

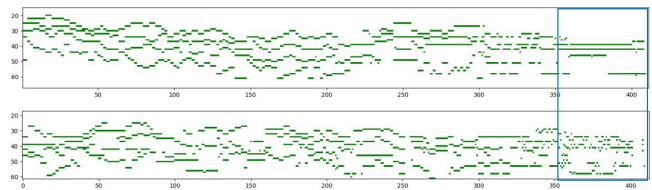

**Fig.7** We compared the Transformer baseline (above) and the Choir Transformer (below) based on the relative position attention mechanism to generate music. The horizontal axis is the number of notes.

We analyzed the generated samples and compared the music generated using the Transformer baseline and relative positional attention mechanism. The pianoroll shows that the model based on the relative position attention can generate a longer effective length. At the same time, the

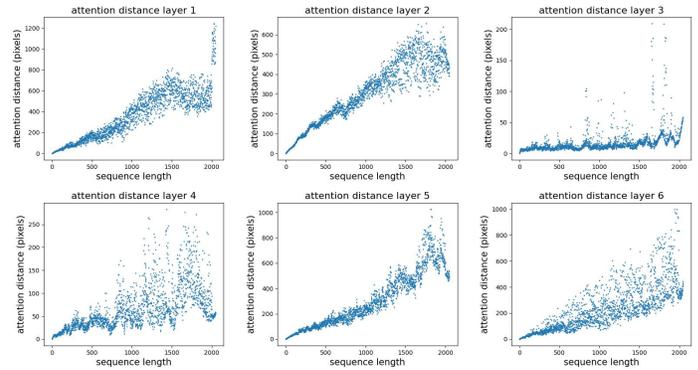

**Fig.8** The attention distance of each layer of Choir Transformer, the first line from left to right is the attention distance from the first layer to the third layer, and the second line is the attention distance from the fourth layer to the sixth layer from left to right.

Transformer baseline is more likely to generate long single notes after the length reaches 350 characters, which reduces the musicality.

We also analyze the attention matrix in each layer of the Choir Transformer encoder. Attention distance was computed by averaging the distance between the query character and all other characters, weighted by the attention weight, as shown in **Fig.8**.

The attention of different layers aggregates information at different scales. The first, second, fifth, and sixth layers of attention aggregate global information, while the third and fourth layers of attention pay more attention to local information within a distance of 200 characters. Each layer's attention is different, reflecting the aggregation ability of Choir Transformer's global and local information, and can generate more accurately based on more information.

### 4.CONCLUSIONS

We propose the Choir Transformer, which can better model polyphonic music and integrate global and local information to generate notes. This paper also proposes a relative position attention mechanism to better adapt to the modeling of musical structures. Experiments showed that the Choir Transformer surpasses every model we know for polyphonic music generation and have better performance on harmonic metrics, closer to the music of Bach. The generated melody and rhythm can be adjusted through input and the model has no limit to the Bach style and can generate different styles of polyphonic music, such as folk music, pop music, and so on. In future work, we will explore how to generate vocal textures that more closely resemble human composition. Due to the training method imposed by teacher-forcing and the cross-entropy loss function, the model is still trained by learning the distribution of polyphonic music. We will explore using a more musical way to generate polyphonic music in the future.